\documentclass[12pt]{article}
\usepackage{rotating}
\usepackage{bm}
\usepackage{multirow}
\usepackage{amsmath, latexsym}
\usepackage{graphicx,psfrag,epsf}
\usepackage{amsfonts}
\usepackage{amssymb}
\usepackage{subfig}
\usepackage{enumerate}
\usepackage{setspace}
\usepackage{epstopdf}
\usepackage[hidelinks]{hyperref} 
\usepackage{pdflscape}
\usepackage{color}
\usepackage{placeins}
\usepackage[round]{natbib}
\usepackage{algorithm}
\usepackage{tkz-graph}  
\usetikzlibrary{calc}
\usepackage{tikz}
\tikzstyle{line} = [draw, -latex']
\usetikzlibrary{arrows.meta}
\usetikzlibrary{shapes.geometric}
\newcommand{\blind}{0}

\newcommand{\norm}[1]{\left\lVert#1\right\rVert}

\newcommand{\given}{\,|\,}

\usepackage{lipsum}
\usepackage{mathtools}

\usepackage{listings}
\begin{document}
\def\spacingset#1{\renewcommand{\baselinestretch}%
{#1}\small\normalsize} \spacingset{1}

\if0\blind
{
  \title{\bf Bayesian State Space Modeling of Physical Processes in Industrial Hygiene}
  \author{Nada Abdalla\hspace{.2cm}\\
    Department of Biostatistics, University of California-Los Angeles.\\
    Sudipto Banerjee \hspace{.2cm}\\
    Department of Biostatistics, University of California-Los Angeles.\\
    Gurumurthy Ramachandran\\
    Department of Environmental Health and Engineering, \\Bloomberg School of Public Health, Johns Hopkins University.\\
    Susan Arnold\\
    Division of Environmental Health Sciences, School of Public Health, \\
    University of Minnesota.
    }
  \maketitle
} \fi

\if1\blind
{
  \bigskip
  \bigskip
  \bigskip
  \begin{center}
    {\LARGE\bf Title}
\end{center}
  \medskip
} \fi

\bigskip
\begin{abstract}
Exposure assessment models are deterministic models derived from physical-chemical laws. In real workplace settings, chemical concentration measurements can be noisy and indirectly measured. In addition, inference on important parameters such as generation and ventilation rates are usually of interest since they are difficult to obtain. In this paper we outline a flexible Bayesian framework for parameter inference and exposure prediction. In particular, we devise Bayesian state space models by discretizing the differential equation models and incorporating information from observed measurements and expert prior knowledge. At each time point, a new measurement is available that contains some noise, so using the physical model and the available measurements, we try to obtain a more accurate state estimate, which can be called filtering. We consider Monte Carlo sampling methods for parameter estimation and inference under nonlinear and non-Gaussian assumptions. The performance of the different methods is studied on computer-simulated and controlled laboratory-generated data. We consider some commonly used exposure models representing different physical hypotheses.
\end{abstract}

\noindent%
{\it Keywords: Bayesian modeling; Eddy-diffusion; Exposure assessment; Industrial hygiene; Kalman filters; Physical Models; State-Space Modeling; Two-zone model; Well-mixed model.} 
\vfill
\hfill {\tiny technometrics tex template (do not remove)}

\newpage
\spacingset{1.45} 

\newpage
\section{Introduction}\label{sec:intro}
In industrial hygiene, estimation of a worker's exposure to chemical concentrations in the workplace is an important concern. In many situations, chemical concentrations are unobserved directly and partial noisy measurements are available. Exposure models aim at capturing the underlying physical processes generating chemical concentrations in the workplace. Exposure modeling through statistical and mathematical models may provide more accurate exposure estimates than monitoring \citep{nicas}. Industrial hygienists seek to infer these latent processes from the available measurements as well as quantification of uncertainty in parameter estimation. For example, generation and ventilation rates are crucial parameters that are difficult to obtain since most workplaces do not collect information routinely. 
Traditional approaches involve using deterministic physical models that ignore the existence of uncertainty by assigning values to those parameters \citep{eddy}. These approaches however don't provide accurate representation in a real workplace environment. Bayesian methods combining professional judgment from experts and direct measurements \citep{gelman} were successful in different settings \citep{vadali}. For example, \cite{bayesih2z} introduced a nonlinear regression on the solution of the differential equations representing the underlying physical model within a Bayesian setting for the two-zone model using Gaussian errors. The model has some limitations since it ignores extraneous factors and variations and requires a closed-form solution of the  differential equations. This severely limits the number of applicable physical models. \cite{b2z} introduced an \texttt{R} package (\b{B2Z}), which implements the Bayesian two-zone model proposed by \cite{bayesih2z}. \cite{bayesih} demonstrated that straightforward Bayesian regression can be ineffective in predicting exposure concentrations in industrial workplaces since the information is limited to partial measurements. They introduced a process-based Bayesian melding approach where measurements are related to the physical model through a stochastic process that captures the bias in the physical model and a measurement error. The resulting inference suffers from inflated variability because of the additional complexities in the model, cumbersome computations and opaque interpretation.

Physical models for industrial hygiene are represented by differential equations that model the rate of change in concentrations. We propose using Bayesian state space models by discretizing the physical model differential equations and incorporating information from observed measurements and experts prior knowledge. This approach will enrich the existing methods, as industrial hygienists will no longer be restricted to fitting a confined selection of physical models amenable to analytic solutions. Any conceivable physical model, in theory, can be accommodated. Neither will they be restricted to Gaussian data, an assumption that most industrial hygiene practitioners will agree is rarely tenable, especially given the small to moderate number of measurements they have to deal with. 

At each time point, a new measurement is available that contains some noise, so using the physical model and the available measurements, we try to obtain a more accurate state estimate, which can be called filtering. The importance of filters lies in their ability to produce estimates of the latent process using information generated by the observations which may provide a poor representation of the latent process if used alone. The aim is to infer the latent process using those observations, along with the physical model that theoretically describes it, as well as incorporating professional knowledge. We consider Monte Carlo based filtering methods for parameter estimation and inference in state space models. We also relax the assumption of Gaussian error terms and consider other alternatives.
 
In particular, we consider different filtering methods under different assumptions. The widely deployed Kalman filter (KF) \citep{kf} offers an optimal solution under linearity and normality assumptions. State-by-state update sampler \citep{mcmcstsp} can provide state estimates under nonlinear and/or non-Gaussian models. 
The different models are compared and assessed using computer-simulated data and lab-generated data. In the lab-generated data, most of the model parameters are known up to a considerable level of accuracy. Experiments were conducted in a controlled chamber that mimics real workplace settings where concentrations were generated at different ventilation and generation rates and under different exposure physical models.

Our contribution in this article expands upon the existing exposure models to allow for better prediction of the quantities of interest. The article is organized as follows. Section~\ref{sec:models} provides a brief review of three families of commonly referenced exposure physical models. Section~\ref{sec:implement} describes the Bayesian approaches used. Section~~\ref{sec:results} illustrates our model through applying it to the simulated data and lab-generated data. Section~\ref{sec:discussion} concludes the article suggesting some future work.

\section{Physical models and their statistical counterparts}\label{sec:models}
Bayesian state space representations for exposure assessment models combine direct measurements of the environmental exposure, physical models and prior information. There are several physical models varying in their level of complexity \citep{ram}. Three commonly used families of physical models are the well-mixed compartment (one-zone) model, the two-zone model and the turbulent eddy diffusion model. We use discrete approximations to the deterministic physical models and introduce stochastic error terms to derive corresponding dynamic statistical models. This obviates the need for exact analytic solutions to the differential equations, which can be sensitive to the choice of initial conditions. Prior specifications for the model parameters produce Bayesian state space models (SSMs). 

Dynamic steady-state models combine measurements with the true underlying state. They are composed of (i) a measurement equation that relates the observations (or some function thereof) to the true concentrations; and (ii) a transition equation describing the concentration change from time $t$ to time $t+\delta_t$. We will derive the dynamic models from the respective differential equations for three popular physical models in industrial hygiene. 

\subsection{Well-mixed compartment (one-zone) model}\label{sec:onecomp}
\begin{figure} [ht]
\centering
\captionsetup{justification=centering,margin=2cm}
\includegraphics[scale=0.5]{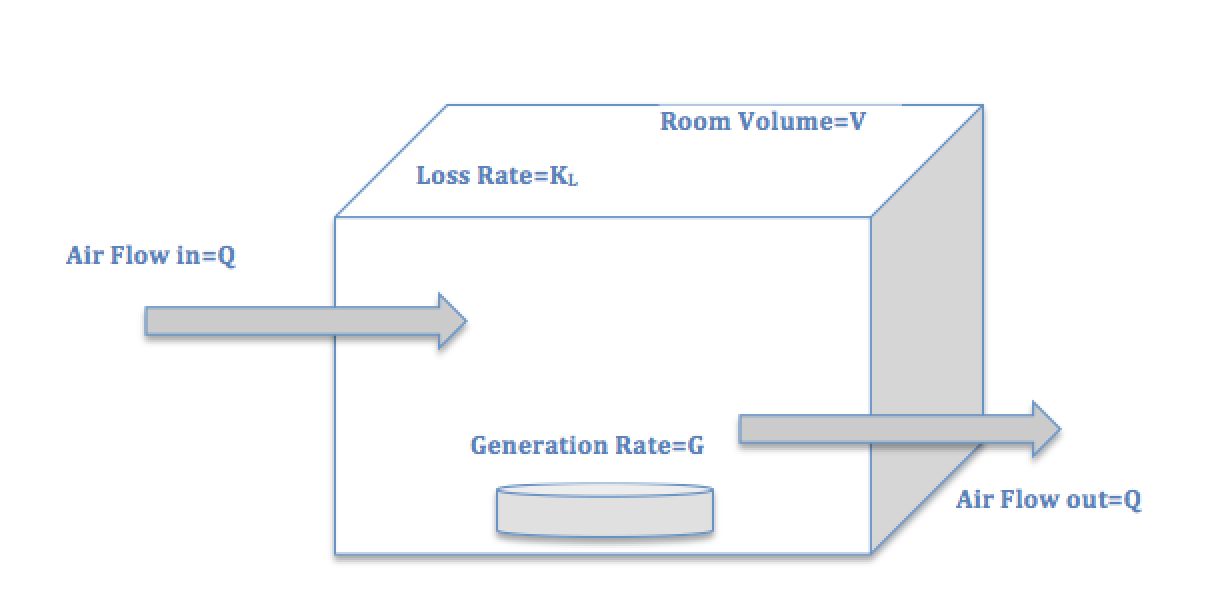}
 \caption{One-zone model schematic showing key model parameters; generation rate $G$, ventilation rate $Q$ and loss rate $K_L$} 
 \label{fig:onezone} 
 \end{figure}

\noindent The well-mixed compartment model assumes that a source is generating a pollutant at a rate $G$ (mg/min) in a room of volume $V$(m$^3$) with ventilation rate $Q$(m$^3$/min). The room is assumed to be perfectly mixed, which means that there is a uniform concentration of the contaminant throughout the room (Figure~\ref{fig:onezone}). The loss term $K_L$(mg/min) measures the loss rate of the contaminant due to other factors such as chemical reactions or the contaminant being absorbed by the room surfaces.
  
The differential equation describing this model is 
\begin{equation}\label{eq:exactdiffone} 
V \frac{d}{dt} C(t)+\left(Q+K_LV\right)C(t)=G \;. 
\end{equation}
The solution to the differential equation is 
\begin{equation}\label{eq:exactone} 
C(t)=\text{exp}\{-t(Q+K_LV)/V\} C(t_0)+\left((Q+K_LV)/V\right)^{-1}\left[1-\text{exp}\{-t(Q+K_LV)/V\}\right]G/V \;. 
\end{equation}
Theoretically, the steady state concentration is the limit of $C(t)$ as $t \rightarrow \infty$ which is $G/Q$  (mg/m$^3$). Details of the steady state solution are provided in the supplementary material. Further specifications yield the Bayesian SSM corresponding to (\ref{eq:exactdiffone}). For example,
\begin{align}\label{eq:discreteone}
&\mbox{Measurement: }\; Z_t = f(C_t) + \nu_t\;, \quad \nu_t \stackrel{iid}{\sim}P_{\nu,\theta_\nu}\;; \nonumber\\
&\mbox{Transition: }\; C_{t+\delta_t} = \left(1-\delta_t\frac{Q+K_LV}{V}\right)C_t +\delta_t\frac{G}{V} + \omega_{t}\;,\quad \omega_t \stackrel{iid}{\sim}P_{\omega, \theta_\omega}\;. \nonumber\\
& Q \sim Unif(a_Q,b_Q)\;;\quad G \sim Unif(a_G,b_G)\;;\quad K_L \sim Unif(a_{K_L},b_{K_L})\;;\quad \sigma^2 \sim IG(a_\sigma,b_\sigma)\;; \quad 
\end{align} 
where $Z_t$ represents measurements (perhaps transformed), $f(\cdot)$ is a function that maps $C_t$ to the scale of $Z_t$, $P_{\nu,\theta_\nu}$ and $P_{\omega,\theta_\omega}$ are probability distributions to be specified, while the prior distributions for the physical parameters are customarily specified as uniform within certain fixed physical bounds.

\subsection{Two-zone model}\label{sec:twocomp}
\begin{figure} [ht]
\centering
\captionsetup{justification=centering,margin=2cm}
\includegraphics[scale=0.5]{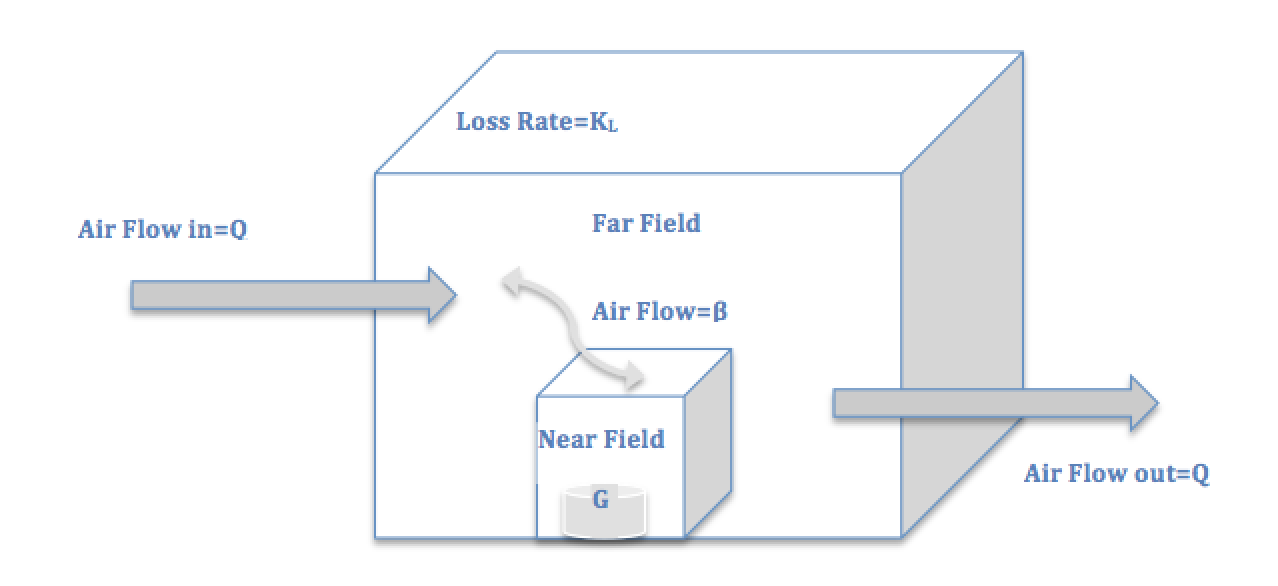}
 \caption{Two-zone model schematic showing key model parameters; 
  generation rate $G$, ventilation rate $Q$, airflow $\beta$ and loss rate $K_L$} 
 \label{fig:2zone} 
 \end{figure}

The two zone model assumes the presence of a source for the contaminant in the workplace. Two zones or regions are defined: (i) the region closer to the source is called the \textit{``near field"}, while the rest of the room is called the far \textit{``far field"}, which completely encloses the near field. Both fields are assumed to be a well-mixed box, i.e., two distinct places that are in the same field have equal levels of concentration of the contaminant. Similar to the one-zone model, this model assumes that a contaminant is generated at a rate $G$(mg/min), in a room with supply and exhaust flow rates (ventilation rate) $Q$(m$^3$/min) and loss rate by other mechanisms $K_L$(mg/m$^3$). This model includes one more parameter that indicates the airflow between the near and the far field $\beta$(m$^3$/min).  The volume in the near field is denoted by $V_N$(m$^3$) and the volume in the far field is denoted by $V_F$(m$^3$). Figure~\ref{fig:2zone} illustrates the dynamics of the system. 
  
The following system of differential equations represents the two-zone model \\
\begin{align}\label{eq:exactdifftwo}\overbrace{ {\LARGE \frac{d}{dt}}\left[ {\begin{array}{c} C_N(t)\\ C_F(t) \\ \end{array} } \right]}^{\frac{d}{dt} C(t)} =\overbrace{ \left[{\begin{array}{cc} -\beta/V_N & \beta/V_N \\
\beta/V_F & -(\beta+Q)/V_F +K_L \end{array} } \right]}^A \overbrace{\left[{\begin{array}{c} C_N(t)\\ C_F(t)\end{array}} \right]}^{C(t)} +\overbrace{\left[{\begin{array}{c} G/V_N\\ 0 \end{array} }\right] }^g\;.
\end{align}
The solution to the differential equations is 
\begin{equation}\label{eq:exacttwo} C(t)=\exp(tA) C(t_0)+A^{-1}\left[\exp(tA)-I\right]g\;, 
\end{equation}
where $\exp(tA)$ is the matrix exponential. Theoretically, for large values of $t$, the steady state concentration in the near field is $G/Q+G/\beta $ (mg/m$^3$), and $G/Q$ (mg/m$^3$) in the far field. We note that the matrix exponential may be numerically unstable to compute in general. For example, for non-diagonalizable matrices a Jordan decomposition \citep[see, e.g.,][]{labanerjee} may be required, which is very sensitive to small perturbations in the elements of $A$. Hence, we will avoid this approach.

Analogous to (\ref{eq:discreteone}), the discrete counterpart of (\ref{eq:exactdifftwo}) can be 
\begin{align*}
& \mbox{Measurement: }\; Z_t = f(C_t) + \nu_t\, ,\; \nu_t \stackrel{iid}{\sim}P_{\nu_t,\theta_\nu}\;; \nonumber \\
&\mbox{Transition: }\; C_{t+\delta_t} = \left(\delta_t A(\theta_c; x)+I\right) C_t + \delta_t g(\theta_c; x)+\omega_{t}\;;\quad \omega_t \stackrel{iid}{\sim}P_{\omega_t, \theta_\omega}\;; \nonumber \\
& Q \sim Unif(a_Q,b_Q)\;;\quad G \sim Unif(a_G,b_G)\;;\quad K_L \sim Unif(a_{K_L},b_{K_L})\;;\quad \beta \sim Unif(a_{\beta}, b_{\beta})\;,
\end{align*}
where $Z_t$ is the $2\times 1$ vector with near-field and far-field measurements (or some function thereof) at time $t$, $C_t$ is the unobserved concentration state at time $t$, $A(\theta_c;x)=\left[{\begin{array}{cc} -\beta/V_N & \beta/V_N \\
\beta/V_F & -(\beta+Q)/V_F +K_L \end{array} } \right] $ and $g(\theta_c;x)=\left[{\begin{array}{c} G/V_N\\ 0 \end{array} }\right]$. 
Similar to the one-zone model, we will specify distributions for $\nu_t$ and for $\omega_t$, where $\theta_\nu$ and $\theta_\omega$ are parameters in $P_{\nu,\theta_\nu}$ and $P_{\omega,\theta_\omega}$, respectively. 

\subsection{Turbulent eddy diffusion model}
\begin{figure} [ht]
\centering
\captionsetup{justification=centering,margin=2cm}
\includegraphics[scale=0.5]{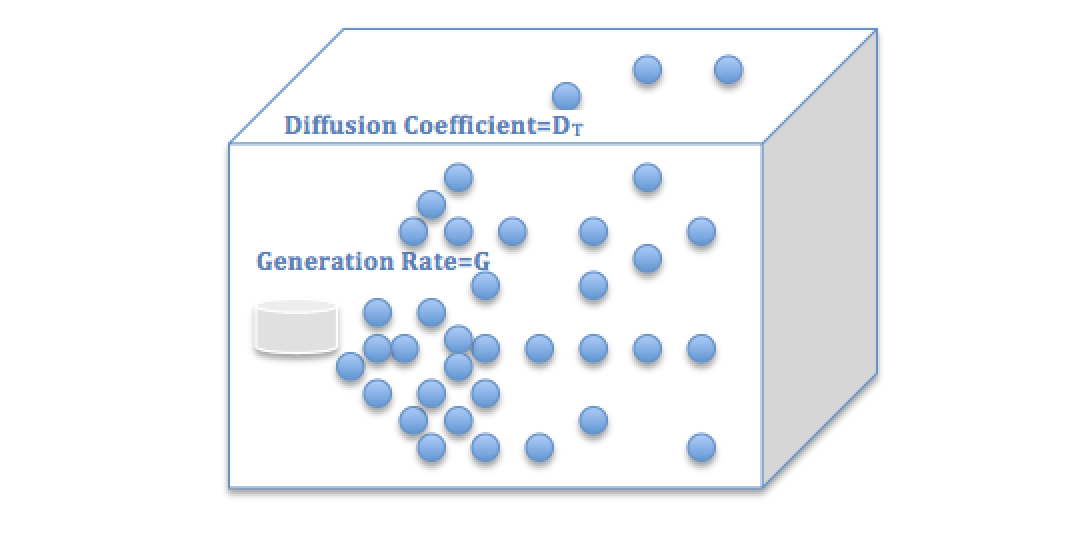}
 \caption{Eddy diffusion model schematic showing key model parameter; diffusion coefficient $D_T$} 
 \label{fig:eddy} 
 \end{figure}

In real workplace settings, the rooms may neither be perfectly mixed nor consist of well-mixed zones. Furthermore, the concentration state could depend upon space and time. A popular model for such settings is the turbulent eddy diffusion model. This model accounts for a continuous concentration gradient from the source outward. It takes into account the worker's location relative to the source. The concentration $C(s,t)$ is a function of the location $s=(x,y)$ in a two-dimensional Euclidean coordinate frame and time $t$. Without loss of generality, the source of the contaminant is assumed to be at coordinate $(0,0)$. The parameter that is unique to this model is the turbulent eddy diffusion coefficient $D_T$(m$^2$/min). It describes how quickly the emission spreads with time (Figure~\ref{fig:eddy}) and is assumed to be constant over space and time. 
There has been very little research on the values of $D_T$ due to the difficulty of measuring it. Some studies suggest a relationship between $D_T$ and air change per hour (ACH) \citep{eddyram}. We will provide inference for this parameter.

The exact contaminant concentration at location $s$ relative to the source of emission is
\begin{equation}\label{eq:eddyexact}
C(s,t)=\frac{G}{2 \pi D_T \norm{s}} \left\{1-\text{erf} \left(\frac{\norm{s}}{\sqrt{4D_Tt}}\right) \right\}, \end{equation}
where $\text{erf}(z)=\frac{2}{\pi} \int_0^z \exp (-u^2)du$. The steady state concentration at location $s$ is theoretically the limit of the concentration as $t \rightarrow \infty$, which is $G/(2 \pi D_T (s))$ (mg/m$^3$).

The following differential equation represents the change in concentration over time
\begin{equation} 
\frac{d}{dt}C(s,t)=\frac{G}{4(D_T \pi t)^{3/2}} \exp \left(-||s||^2/4D_T t\right).\nonumber 
\end{equation} 
A general dynamic modeling framework accounting for space and time is as follows: 
\begin{align}\label{eq: eddy_discrete}
&\mbox{Measurement: }\; Z(t,s) = f(C(t,s)) + \nu_t(s)+\eta_t\,, \; \nu_t(s) {\sim} P_{\nu_t(s),\theta_\nu}\,, \; \eta_t \sim P_{\eta_t, \theta_\eta}\;;\nonumber\\
&\mbox{Transition: }\; C(s,t+\delta_t) = C(s,t)+\delta_t \frac{G}{4(D_T \pi t)^{3/2}} \exp \left(-||s||^2/4D_T t\right)+\omega(s,t+\delta_t)\,,\; \omega(s,t) {\sim}P_{\omega_{t,s},\theta_\omega}\;; \nonumber \\
& D_T \sim Unif (a_{D_T}, b_{D_T})\;;\quad G \sim Unif(a_G, b_G)\; ,
\end{align}
where $P_{\nu_t(s),\theta_\nu}$ and $P_{\omega_{t,s},\theta_\omega}$ are spatial-temporal stochastic processes. Note that $\nu_t(s)$ is a spatial-temporal process discrete in time and continuous in space. This is reasonable because the measurments are taken over discrete time intervals and the estimation for the latent concentration states are required at those intervals. On the other hand, $\omega(s,t)$ would ideally be a process continuous in both space and time because it models spatial-temporal associations between concentration states at arbitrary space-time coordinates. 

\section{Model Implementation and Assessment}\label{sec:implement}

\noindent For each physical model in Section~\ref{sec:models} we will consider two different Bayesian SSMs. We will refer to the first as a Gaussian SSM. Gaussian (linear) SSMs result from specifying $f(C_t) = B_tC_t$, where $B_t$ is a known $p\times p$ design matrix (usually the identity matrix), $P_{\nu,\theta_\nu} \equiv N(0,\Sigma_{\nu})$ and $P_{\omega,\theta_\omega} \equiv N(0,\Sigma_{\omega})$ are $p$-variate Gaussian densities. These deliver accessible distribution theory for updating parameters using Kalman-filters or Gibbs samplers. Let ${\cal T} = \{t_1,\ldots, t_n\}$ be timepoints where concentration measurements $Z_t$ have been measured. A Bayesian hierarchical SSM is 
\begin{align}\label{eq: bayesian_Gaussian_SSM}
& p(\theta_c) \times IW(\Sigma_{\omega}\given r_{\omega}, S_{\omega}) \times IW(\Sigma_{\nu}\given r_{\nu}, S_{\nu}) \times N(C_{t_0}\given m_0, \Sigma_0) \nonumber \\ 
& \qquad \qquad  \times \prod_{i=1}^n N(C_{t_i}\given A_{t_i}(\theta_c)C_{t_{i-1}} + \delta_i g_{t_i}, \Sigma_{\omega}) \times \prod_{i=1}^{n} N(Z_{t_i}\given B_{t_i}C_{t_i}, \Sigma_{\nu})\; ,
\end{align}
where $p(\theta_c)$ is the prior distribution on $\theta_c$, $\delta_i = t_{i} - t_{i-1}$, and the other distributions follow definitions as in \cite{gelman}. Gibbs updates are implemented using $p(C_{t_i}\given \cdot) = N(C_{t_{i}}\given M_{t_i}m_{t_i}, M_{t_i})$, where $m_{t_i}=\Sigma_{\nu}^{-1}Z_{t_i}+\Sigma_{t_i|t_{i-1}}^{-1}A_{t_i}(\theta_c)C_{t_{i-1}}$ and $M_{t_i}=(\Sigma_\nu^{-1}+\Sigma_{t_i|t_{i-1}}^{-1})^{-1}$, where $\Sigma_{t_i|t_{i-1}} = A_{t_i}(\theta_c) M_{t_{i-1}}A_{t_i}(\theta_c)^T+\Sigma_\omega$ and $M_{t_0}=\Sigma_0$, $p(\Sigma_{\nu}\given \cdot) = IW(\Sigma_{\nu}\given r_{\nu|\cdot}, S_{\nu|\cdot})$ and $p(\Sigma_{\omega}\given \cdot) = IW(\Sigma_{\omega}\given r_{\omega|\cdot}, S_{\omega|\cdot})$, where $r_{\nu|\cdot} = r_\nu+n$, $S_{\nu|\cdot}=S_\nu+\sum_{i=1}^n(Z_{t_i}-B_{t_i}C_{t_i})(Z_{t_i}-B_{t_i}C_{t_i})^T$, $r_{\omega|\cdot}=r_\omega+n$ and $S_{\omega|\cdot}=S_\omega+\sum_{i=1}^n(C_{t_i}-A_{t_i}(\theta_c)C_{t_{i-1}})(C_{t_i}-A_{t_i}(\theta_c)C_{t_{i-1}})^T$. 

Note that the two-zone model has $p=2$, while the one-compartment and eddy-diffusion models have $p=1$. Gaussian Bayesian SSMs for $p=1$ specify   
$P_{\nu,\theta_\nu} \equiv N(0,\sigma^2)$ and $P_{\omega,\theta_\omega} \equiv N(0,\tau^2)$. The measurement equation is linear in the state $C_t$. The $IW(\cdot,\cdot)$ priors in (\ref{eq: bayesian_Gaussian_SSM}) are replaced by $IG(\sigma^2\given a_{\sigma}, b_{\sigma})$ and $IG(\tau^2\given a_{\tau}, b_{\tau})$. The full conditionals now assume the form $p(C_{t_i}\given \cdot) = N(C_{t_{i}}\given M_{t_i}m_{t_i}, M_{t_i})$, where $m_{t_i}=\sigma^{-2}Z_{t_i}+\sigma_{t_i|t_{i-1}}^{-2}A_{t_i}(\theta_c)C_{t_{i-1}}$ and $M_{t_i}=1/(\sigma^{-2}+\sigma_{t_i|t_{i-1}}^{-2})$, where $\sigma_{t_i|t_{i-1}}^{2}=A_{t_i}(\theta_c)^2 M_{t_{i-1}}+\tau^2$, $p(\sigma^2\given \cdot) = IG(\sigma^2\given a_{\sigma|\cdot}, b_{\sigma|\cdot})$ and $p(\tau^2\given \cdot) = IG(\tau^2\given a_{\tau|\cdot}, b_{\tau|\cdot})$, where $a_{\sigma|\cdot} = a_{\sigma}+n/2$, $b_{\sigma|\cdot}=b_{\sigma}+\sum_{i=1}^n(Z_{t_i}-B_{t_i}C_{t_i})^2/2$, $a_{\tau|\cdot}=a_{\tau}+n/2$ and $b_{\tau|\cdot}=b_{\tau}+\sum_{i=1}^n(C_{t_i}-A_{t_i}(\theta_c)C_{t_{i-1}})^2/2$. 

Although Gausian SSMs are very popular in dynamic modeling of physical systems, especially due to convenient updating schemes, the Gaussian assumption for the concentration measurements may be untenable. Our second Bayesian SSM assumes that $Z_t = \log Y_t$ are log-concentration measurements and $f(C_t) = \log C_t$ in the measurement equation. We still specify $P_{\nu,\theta_\nu}$ as Gaussian, which means that $Z_t$'s are log-normal and is probably a more plausible assumption than in Gaussian SSMs. In the transition equation, again the Gaussian assumption on $\omega_t$ seems implausible: if the measurements of the state are log-normal, then why should $C_t$ be Gaussian? Since $C_t$ is positive, a Gamma or log-normal specification for $P_{\omega,\theta_\omega}$ seems much more plausible. For $p=2$, we will specify logarithmic bivariate normal distributions, while for $p=1$ we will explore with both Gamma and log-normal densities. We will refer to all of these models as non-Gaussian Bayesian SSMs.     

The turbulent eddy-diffusion model requires some further specifications. While the framework in (\ref{eq: eddy_discrete}) is rich, unfortunately it will not usually be applicable to practical industrial hygiene settings because typically very few measurements are available over distinct locations in a workplace chamber and estimating the processes will be unfeasible. Hence, we will need simpler specifications. For example, we can consider a setting with locations $\{s_1, s_2,\ldots, s_m\}$ 
and $n$ time-points. We fit the model in (\ref{eq: eddy_discrete}) with $Z_t(s_i) = \log Y_t(s_i)$ are log-concentration measurements and $f(C_t(s_i)) = \log C_t(s_i)$. We further specify $P_{\eta_t, \theta_\eta}$ as a white-noise process, i.e., $\eta_t \stackrel{iid}{\sim} N(0,\tau^2)$ for every $t$ and $s$, and $P_{\nu_t(s),\theta_\nu}$ is a temporally indexed spatial Gaussian process with an exponential covariance function, independent across time. This means that the $m\times 1$ vector $\nu_t \stackrel{ind}{\sim} N(0,\sigma_t^2 R_t(\phi_t))$, where $R_t(\phi_t)$ is an $m\times m$ matrix with $(i,j)$-th element $\exp(-\phi_t d_{ij})$ and $d_{ij} = \|s_i-s_j\|$.

Note that $P_{\nu_t(s),\theta_\nu}$ can, in theory, be a continuous-time spatial-temporal process specified through a space-time covariance function \citep[see, e.g.,][]{banerjee}. Alternatively, one could treat time as discrete and evolving, for each location $s$, as an autoregressive process so that $\nu_t(s) = \gamma \nu_{t-1}(s) + \eta_t(s)$ with $\eta_t(s)$ being spatial processes independent across time \citep[see, e.g.,][]{wikleandcressie,GBG}. One could continue to embellish the model in (\ref{eq: eddy_discrete}) using spatial-temporal structures that represent richer hypotheses and more flexible modeling. However, in realistic industrial hygiene applications such specifications will rarely lead to estimable models given the scarcity of data points. For example, most settings will provide measurements from only a handful of locations (e.g., $m \sim 5$) and some moderate numbers of time points (e.g., $n \sim 100$). Therefore, we will not explore these specifications any further. Moreover, even when we assume independence across time it will be difficult to estimate models with time-varying spatial process parameters. Hence, we let $\nu_t \stackrel{iid}{\sim} N(0,\sigma^2 R(\phi))$ so that each $m\times 1$ vector $\nu_t$ has the same $m$-variate Gaussian distribution.    

Finally, we turn to smoothing and filtering. Smoothing is achieved by evaluating at each time point $t_i$ the posterior expectation of the concentration value given the entire observed data $y = \{y_{t_i} : i=1,2,\ldots,n\}$, including observations before and after $t_i$. Thus, we sample from the posterior density $p(C_{t_i}\given y)$ in posterior predictive fashion by sampling a $C_{t_i}$ from its full conditional, $p(C_{t_i}\given \cdot)$, for each sampled value of the parameters. For linear Gaussian SSM, Kalman smoother can be used where the smoothed distribution  at time $t$ also follows a Gaussian distribution. For the nonlinear non-Gaussian SSM, \cite{smoothingalg} provided a discussion of the different smoothing approaches. This provides an idea about the structure of the smoothing distribution of the collection of states \citep{MCsmoothing}. Filtering, on the other hand, aims to estimate the posterior expectation of the concentration value $C_{t_{i}}$, given the data up to $t_i$, i.e., $\{y(t_j) : j = 1,2,\ldots,i\}$. We have implemented both smoothing and filtering for all the physical models considered above.

To compare between models, we adopt a posterior predictive loss approach (see, e.g., \cite{D=G+P}). We generate the posterior predictive distributions for each data point, $y_{rep,i}$ for $i=1,2,\ldots,n$ by sampling from $\displaystyle p(y_{rep}\given y) = \int p(y_{rep}\given \theta, \{C_t\}) p(\theta, \{C_t\}\given y) d\theta$, where $\theta$ denotes the full collection of unknown parameters and $\{C_t\}$ is the collection of latent concentrations over the entire time frame. We will compute the posterior predictive mean, $\mu_{rep,i} = \mbox{E}[y_{rep,i}\given y]$, and dispersion, $\Sigma_{rep,i} = \mbox{var}[y_{rep,i} \given y]$, for each $y_{rep,i}$; these are easily calculated from the posterior samples for each $y_{rep,i}$. We will prefer models that will perform well under a decision-theoretic \emph{balanced loss function}  that penalizes departure of replicated means from the corresponding observed values (lack of fit), as well as the uncertainty in the replicated data. Using a squared error loss function, the measures for these two criteria are evaluated as $G = \sum_{i=1}^{n}\|y_i - \mu_{rep,i}\|^2$, where $\|\cdot\|$ denotes the Euclidean norm, and $P = \sum_{i=1}^n \mbox{Tr}(\Sigma_{rep,i})$, where $\mbox{Tr}(A)$ denotes the trace of the matrix $A$. We will use the score $D = G + P$ as a model selection criteria, with lower values of $D$ indicating better models.
 
\section{Data Analysis}\label{sec:results}
In this section we evaluate the performance of the models discussed in Section~\ref{sec:implement}, for the three physical exposure models illustrated in Section~\ref{sec:models}, using computer-simulated datasets as well as experimental lab-generated data. In particular, we consider two models: a Gaussian linear model and a non-Gaussian nonlinear model, and they will be referred to as Gaussian SSM and  non-Gaussian SSM respectively. The prior settings are based on physical knowledge and experience, and discussed in the following section.

The computer-simulated data was generated using R computing environment. The lab-generated data experiments were conducted in test chambers. \cite{arnold} examined parts of this data using the deterministic one-zone and two-zone models and showed that performance is highly reliable on the model assumptions and knowing the generation $(G)$ and ventilation $(Q)$ rates. \cite{eddyram} studied the eddy diffusion data using a deterministic model and concluded that it is suitable for indoor spaces with persistent directional flow toward a wall boundary, as well as in rooms where the airflow is solely driven by mechanical ventilation (no natural ventilation involved). These results imply the need for a more flexible model that accounts for uncertainty and also be used for parameter inference. 

\subsection{Prior settings}\label{sec:priors}
In Bayesian exposure models, reasonable informative priors are usually used, based on expert knowledge and physical considerations \citep{bayesih}.  
We assigned informative priors on the generation rate $G$, ventilation rate $Q$, loss rate $K_L$, airflow rate $\beta$ and diffusion coefficient $D_T$ using uniform distributions for the plausible values of the parameters. For the simulation data, uniform priors were assigned within at least $20 \%$ of the true values following the prior settings in \cite{b2z}. The model parameters used to generate the one-zone model data were taken from physical considerations as illustrated by \cite{bayesih2z} at values, $Q=13.8$ m$^3$/min, $G=351.5$ mg/min, $V=3.8$ m$^3$, $K_L=0.1$ mg/min and $\sigma=0.1$. In the two-zone model, following \cite{bayesih2z}, the generation and ventilation rates were fixed at the same values as in the one-zone model. In addition, $\beta$ was fixed at $5$ m$^3$/min, $V_N=\pi \times 10^{-3}$ m$^3$, $V_F=3.8$ m$^3$, and $\Sigma_\nu=
  \left[ {\begin{array}{cc}
   0.1 & 0 \\
   0 & 0.1 \\
  \end{array} } \right]$. For the eddy diffusion data, we fixed $G=351.5$ mg/min, $D_t=1$ m$^2$/min, $\sigma_\eta^2=0.1$ and used a geostatistical exponential covariance with $\sigma=\phi=1$. 
  
In the one-zone and two-zone models, we assume that $G\sim Unif(281, 482)$, $Q \sim Unif(11,17)$, $K_L\sim Unif(0,1)$, and $\beta \sim Unif(0,10)$ in the two-zone model and $D_T \sim Unif(0,3)$ in the eddy diffusion model. For the exponential covariance function, the spatial range is given by approximately $3/\phi$ which is the distance where the correlation drops below $0.05$. The prior on $\phi \sim Uni(0.5,3)$ implies that the effective spatial range, i.e., the distance beyond which spatial correlation is negligible, is between $1$ and $6$ units.

Wider ranges were considered in the lab-generated data analysis because the exact true values for some of the parameters were unknown but rather a range. The ranges of the true values in the well mixed compartment and two-zone models for $G$, $Q$, $K_L$ and $\beta$ are $(40- 120)$(mg/min), $(0.04- 0.77)$(m$^3$/min), $<0.01$ and $(0.24-1.24)$(m$^3$/min) respectively. We assume that $G\sim Unif(30, 150)$, $Q \sim Unif(0,1)$, $K_L\sim Unif(0,1)$ in the one-zone and two-zone models and $\beta \sim Unif(0,5)$ in the two-zone model. For the eddy diffusion model, the true value for $G$ is $1318$ (mg/sec) and from literature \citep{eddyram} the range for $D_T$ is (0.001-0.2) m$^2$/sec, hence we assigned priors of $G\sim Unif(1104,1650)$ and $D_t \sim Unif(0,1)$.
Non informative priors were assigned to the variance covariance matrices using $IW(3,I)$ \citep{gelman}.

\subsection{Simulation results}\label{sec:simresults}
Monte Carlo filtering methods were used to estimate the latent processes and the model parameters. The effectiveness of the model is assessed through checking whether the 95$\%$ C.I.s of the parameters include the true values, MSE and posterior predictive loss (D=G+P), in addition to graphical assessment.    
\subsubsection{One-zone model}
We simulated 100 exposure concentrations at equally spaced time points using the exact solution to the ODE in equation (\ref{eq:exactone}). The initial concentration $C(0)$ was assigned a value of $1$ mg/m$^3$. Theoretically, the steady state concentration is $G/Q \approx$ 25 mg/m$^3$. 
The models applied to the synthetic data and compared are: Gaussian SSM and  non-Gaussian SSM. The Gaussian SSM in (\ref{eq: bayesian_Gaussian_SSM}) assumes linearity and Gaussian errors, where the Kalman filter equations are used, where 
\begin{align}
A_{t}(\theta_c)=\left(1-\delta_t \frac{Q+K_LV}{V}\right)\quad \text{and} \quad g=\delta_t \frac{G}{V}. \nonumber 
\end{align}

\begin{figure}[h!]
\centering
\captionsetup{justification=centering}
\hspace*{-0.35in}
\includegraphics[scale=0.7]{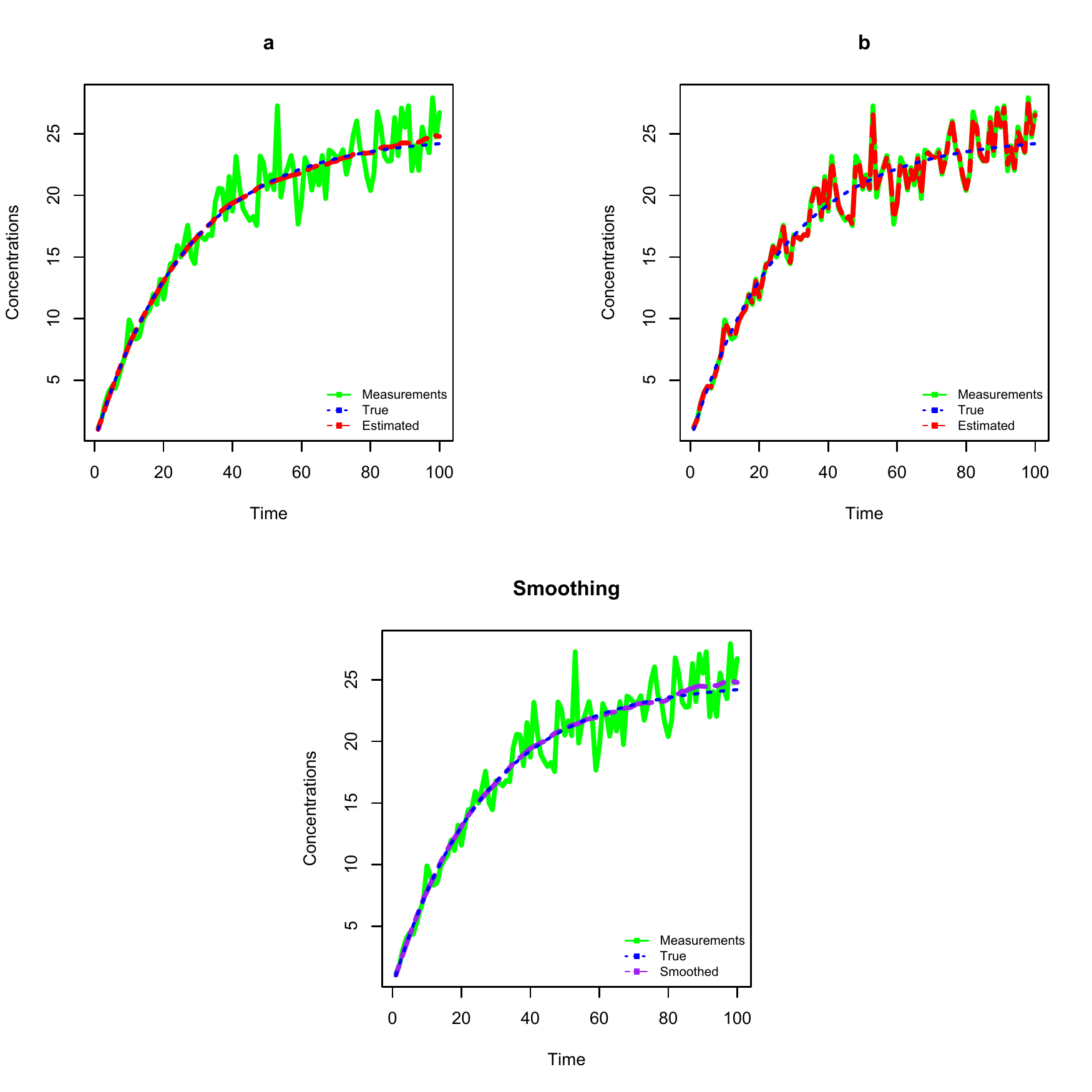}
 \caption{Plot of the simulated concentrations, measurements and the mean of the posterior samples of the latent states conditional on the measurements for:\\ a: Non-Gaussian SSM and b: Gaussian SSM} 
 \label{fig: Model1 post sim mcmc} 
 \end{figure}

Table~\ref{table: Model1 post sim mcmc} shows the medians and 95$\%$ credible intervals of the MCMC posterior samples of the model parameters, MSE and D=G+P for the two aforementioned models. Figure~\ref{fig: Model1 post sim mcmc} shows the simulated concentrations, measurements and the mean of the posterior samples of the latent states conditional on the measurements, in addition to smoothed estimates obtained from the Non-Gaussian SSM filtered states.
Details of the performances are as follows:\begin{itemize}
\item Non-Gaussian SSM: The $95\%$ C.I.s include the true values for all the parameters except $K_L$. The latent state estimates are very close to the true simulated values as shown in Figure~\ref{fig: Model1 post sim mcmc}. 
\item Gaussian SSM:  The $95\%$ C.I.s for the generation rate $G$ and the ventilation rate $Q$ include the true values. The interval for the loss rate $K_L$ does not cover the true parameter value. The model estimates for the latent states are closer to the observed values than the true values. 

\end{itemize}
The D=G+P scores and MSE results suggest that the nonlinear non-Gaussian model outperforms the linear Gaussian one, which is also confirmed in Figure~\ref{fig: Model1 post sim mcmc}. 

\begin{table}[h!]
\small
\caption{Posterior predictive loss (D=G+P), MSE, medians and $95\%$ C.I. of the posterior samples of the one-zone model parameters for the simulated data}
\begin{center}
\setlength\tabcolsep{4pt} 
\begin{tabular}{lccccc}
  \hline 
\hline
 Parameter& Non-Gaussian SSM      & Gaussian SSM\\
\hline
$G(351.5)$ &326.8 (283.3, 351.7)& 363.5(314.2,413.8)\\
$Q(13.8)$ &12.9(11.1, 14.8) & 12.8(11.4, 14.3)\\
$K_L(0.1)$&0.34(0.19,0.78)&0.30(0.28, 0.41)\\
\hline
\hline
D=G+P& 312.2=5.9+306.3& 435.8=232.8+203.0 \\
MSE& 0.07& 2.3\\
 \hline
   \hline
\end{tabular}
\end{center}
\label{table: Model1 post sim mcmc}
\end{table}

\subsubsection{Two-zone model}
We simulated 100 exposure concentrations at the near and far fields at equally spaced time points using the exact solution (\ref{eq:exacttwo}). The initial concentrations $C_N(0)$ and $C_F(0)$ were assigned values $0$ and $0.5$ mg/m$^3$ respectively. Theoretically, the steady state concentration at the near field is $G/Q+G/\beta \approx$ 95 mg/m$^3$, and $G/Q  \approx$ 25 mg/m$^3$ at the far field. 
The Gaussian SSM in (\ref{eq: bayesian_Gaussian_SSM}) assumes linearity and Gaussian errors, such that \begin{align}
A_{t}(\theta_c)=\delta_t A+ I\quad \text{and} \quad g=\delta_t g  . \nonumber \end{align} 
  
\begin{figure}[h!]
\centering
\captionsetup{justification=centering}
\hspace*{-0.35in}
\includegraphics[scale=0.8]{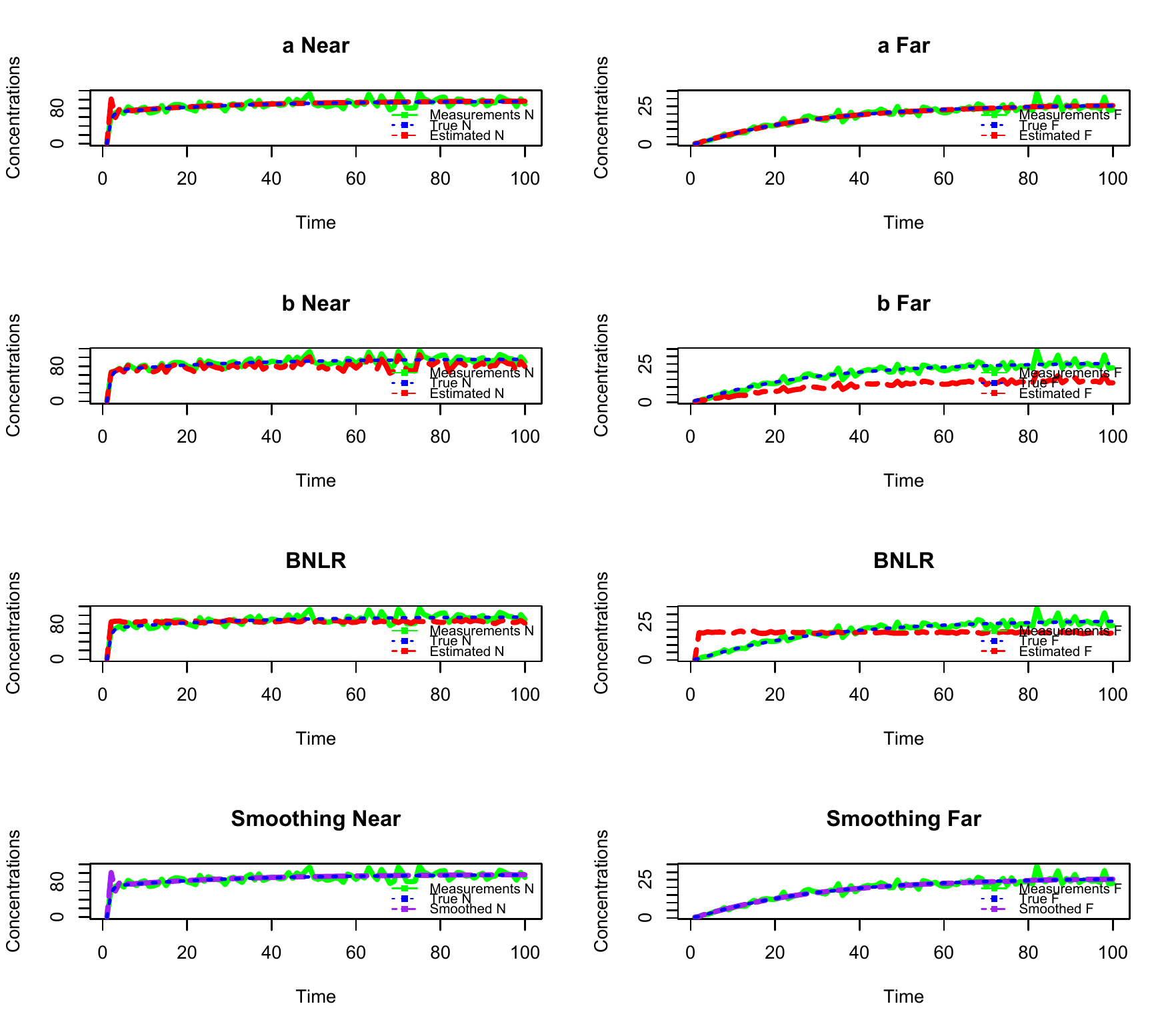}
 \caption{Plot of the simulated near and far fields concentrations, measurements and the mean of the posterior samples of the latent states conditional on the measurements for:\\ a: Non-Gaussian SSM, b: Gaussian SSM and BNLR} 
 \label{fig: Model2 post sim mcmc} 
 \end{figure}

Table~\ref{table: Model2 post sim mcmc} shows the medians and 95$\%$C.Is of the MCMC posterior samples of the model parameters, MSE and D=G+P scores. Figure~\ref{fig: Model2 post sim mcmc} shows the simulated concentrations, measurements and the mean of the posterior samples of the latent states conditional on the measurements at the near and the far fields in addition to smoothed estimates obtained from the non-Gaussian SSM filtered states. Moreover, we compared the performance of the two SSMs to the simple Bayesian nonlinear regression model (BNLR) proposed by \cite{bayesih2z}.
Details of the performances of the three models are as follows:\begin{itemize}
\item Non-Gaussian SSM: The $95\%$ C.I.s include the true values for all the parameters. The estimates of the latent states are close to the true values at both the near field and the far field as shown in Figure~\ref{fig: Model2 post sim mcmc}. 
\item Gaussian SSM:  The $95\%$ C.I.s for all the parameters except the ventilation rate $Q$ do not include the true values. The model estimates of the latent states are closer to the true values at the near field than the far field. 
\item BNLR: The $95\%$ C.I.s include the true values for all the parameters.
\end{itemize}
The D=G+P scores indicate that the non-Gaussian model provides better fit than the BNLR and the Gaussian models. MSE and Figure~\ref{fig: Model2 post sim mcmc} confirm these results.
  \begin{table}[h!]
\small
\caption{Posterior predictive loss (D=G+P), MSE, medians and $95\%$ C.I. of the posterior samples of the two-zone model parameters for the simulated data}
\begin{center}
\setlength\tabcolsep{7pt} 
\begin{tabular}{lcccc}
  \hline 
\hline
 Parameter& Non-Gaussian SSM   &  Gaussian SSM & BNLR \\
\hline
$G(351.5)$&347.3(315.6,379.3)&450.5(395.2, 480.2) & 335.1(302.5,382.6)\\
$Q(13.8)$ &14.7(12.1,16.8)&13.5(11.1, 16.7)&	14.4(11.2, 15.8)\\
$K_L(0.1)$&0.38(0.02,0.78)&0.22(0.16,0.35)&-\\
$\beta(5)$ &5.0(4.3,5.8) & 0.40(0.23,1.2)  & 5.1(4.0, 6.8)	\\
\hline
\hline
\multirow{2}{4em}{D=G+P} 
&1049840= & 1118550=&2504429=\\
&1010905+38934.0&1033428+85121.7&1359016+ 1145413\\
MSE & 15.3& 116.1&54.9\\
 \hline
   \hline
\end{tabular}
\end{center}
\label{table: Model2 post sim mcmc}
\end{table}

\subsubsection{Turbulent eddy diffusion model}
\begin{figure} [h!]
\centering
\captionsetup{justification=centering}
\hspace*{-0.35in}
\includegraphics[scale=0.8]{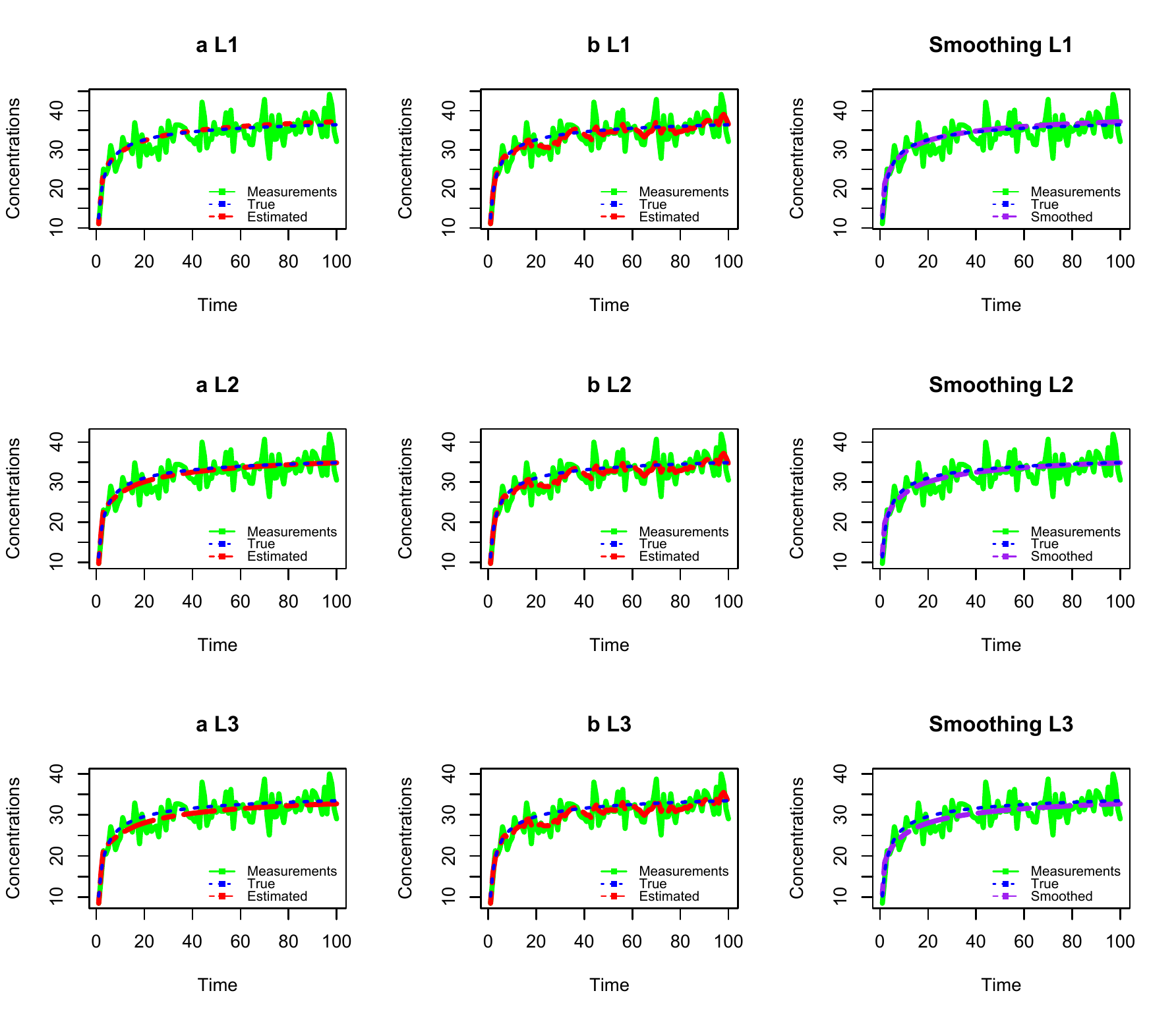}
 \caption{Plot of the simulated concentrations, measurements and the mean of the posterior samples of the latent states conditional on the measurements at three locations for:\\ a: Non-Gaussian SSM and b: Gaussian SSM} 
 \label{fig: Model3 post sim} 
 \end{figure}

\begin{figure} [h!]
\centering
\captionsetup{justification=centering}
\hspace*{-0.5in}
\includegraphics[scale=0.15]{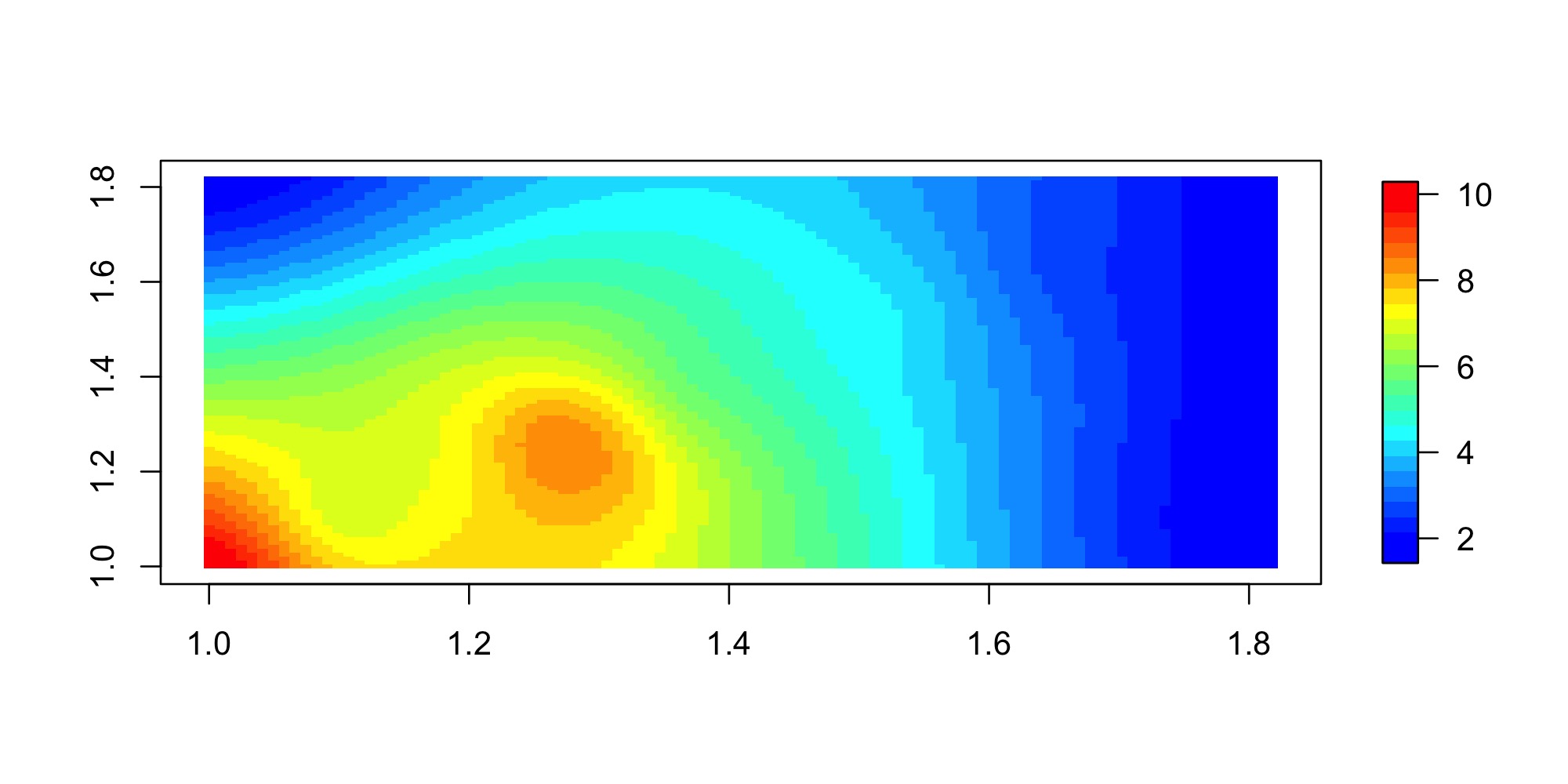}
 \caption{Interpolated surface of the mean of the random spatial effects posterior distribution} 
 \label{fig:spatial} 
 \end{figure}

We simulated a total of 500 exposure concentrations at 5 different locations over equally spaced 100 time points using the exact equation (\ref{eq:eddyexact}). 
Table~\ref{table: Model3 post sim} shows the medians and 95$\%$ C.I.s of the MCMC posterior samples of the model parameters, MSE and D=G+P. Figure~\ref{fig: Model3 post sim} shows the simulated concentrations, measurements and the mean of the posterior samples of the latent states conditional on the measurements at three locations and the smoothed estimates obtained from the non-Gaussian SSM filtered states. Figure~\ref{fig:spatial} shows image plot of the posterior mean surface of the latent spatial process $\nu_t(s)$. The plot indicates higher concentration values near the source of emission at the bottom-left corner and lower values away from the source.
Details of the performance of the two models are as follows:\begin{itemize}
\item Non-Gaussian SSM: The $95\%$ C.I.s include the true values for all the parameters. The estimates of the latent states are close to the true values at the five locations. 
\item Gaussian SSM: The $95\%$ C.I.s include the true value for the generation rate $G$ but not for the eddy diffusion coefficient $D_T$. The model estimates for the latent states are closer to the observed values than the true values. 
\end{itemize}
MSE and D=G+P for the Non-Gaussian SSM indicate a better fit. 
\begin{table}[h!]
\small
\caption{Posterior predictive loss (D=G+P), MSE, medians and 95\% C.I of the posterior samples of the turbulent eddy diffusion model parameters for the simulated data}
\begin{center}
\setlength\tabcolsep{7pt} 
\begin{tabular}{lccc}
  \hline 
\hline
 Parameter& Non-Gaussian SSM &  Gaussian SSM  \\
\hline
$G(351.5)$ &355.9(284.0,477.5)&449.6(301.0,480.5)\\
$D_T(1)$ &1.2(0.9,1.5)&1.4(1.3,1.6)\\
\hline
\hline
D=G+P &7062.4=1564.5+5497.9&22025.7=1112.5+20913.1\\
MSE&3.11&5.55\\
 \hline
   \hline
\end{tabular}
\end{center}
\label{table: Model3 post sim}
\end{table}
\subsection{Experimental Chamber Data Results}\label{sec:dataresults}
In this section we study the performance of the non-Gaussian and Gaussian SSMs on controlled lab-generated data in which solvent concentrations have been measured under different scenarios. We are interested in the inference through the posterior distributions of the parameters $Q$ and $G$ in the one-zone model, in addition to $\beta$ in the two-zone model, and $Q$ and $D_T$ in the eddy diffusion model. 
 
\subsubsection{One-zone model}
A series of studies were conducted in an exposure chamber under different controlled conditions. \cite{arnold} constructed a chamber of size $(2.0 \text{m} \times 2.8  \text{m} \times 2.1  \text{m}=11.8  \text{m}^3)$, where two industrial solvents (acetone and toluene) were released using different generation $G(\text{mg}/\text{min})$ and ventilation $Q( \text{m}^3/ \text{min})$ rates. In particular, three levels of ventilation rates corresponding to ranges of 0.04-0.07 $ \text{m}^3/ \text{min}$, 0.23-0.27 $ \text{m}^3/ \text{min}$ and 0.47-0.77 $ \text{m}^3/ \text{min}$ were used. The loss rate $K_L$ was determined from empirical studies to be $<0.01$. Solvent concentrations were measured every 1.5 minutes. Details of the experiments can be found in \citep{arnold}.

\begin{figure}[h!]
\centering
\captionsetup{justification=centering}
\hspace*{-0.35in}
\includegraphics[scale=0.8]{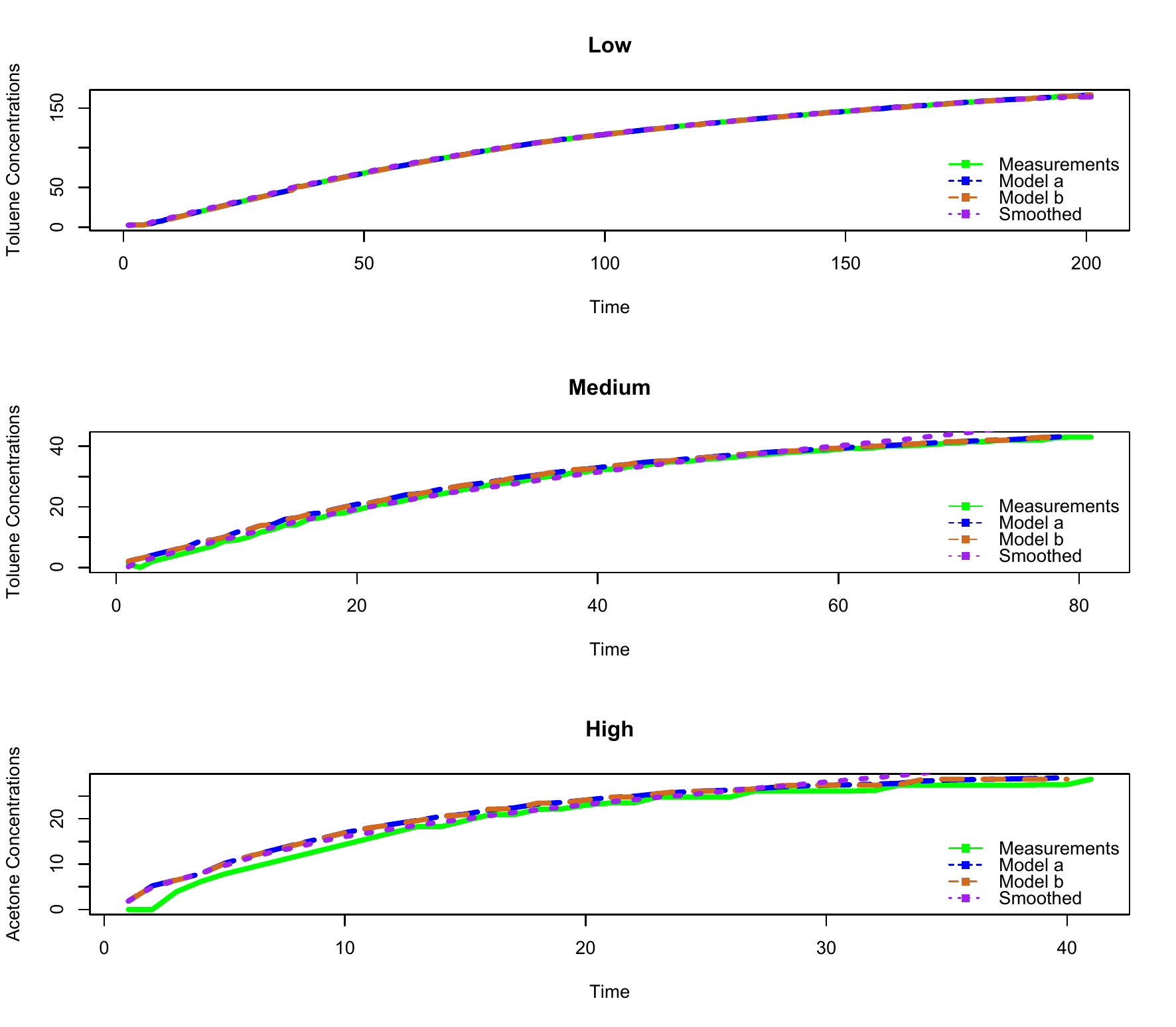}
 \caption{Plot of the measured concentrations and the mean of the posterior samples of the latent states conditional on the measurements for:\\ a: Non-Gaussian SSM and b: Gaussian SSM} 
 \label{fig: Model1 post data} 
 \end{figure}

Table~\ref{table: Model1 post data} shows the medians and 95$\%$ C.I.s of the MCMC posterior samples in addition to MSE and D=G+P. The non-Gaussian SSM 95$\%$ C.I.s cover the true values for both $G$ and $Q$, while Gaussian SSM 95$\%$ C.I.s include the true values for $G$ at low and high ventilation levels. Figure~\ref{fig: Model1 post data} shows that the estimated latent concentrations are close to the measurements. Posterior predictive loss (D=G+P) indicates better fit of the non-Gaussian SSM model.
\begin{table}[h!]
\small
\caption{Posterior predictive loss (D=G+P), MSE, medians and 95\% C.I. of the posterior samples of the one-zone model parameters using toluene and acetone solvents}
\begin{center}
\setlength\tabcolsep{7pt} 
\begin{tabular}{lccccc}
  \hline 
\hline
Parameter &Ventilation level &True value& Non-Gaussian SSM  & Gaussian SSM  \\
\hline
\multirow{3}{4em}{$G$} 
& low&43.2&38.1(30.2,62.9)  &35.3(30.2, 46.7)\\
&medium&43.2&45.06(30.5,101.9)&72.9(45.6,94.9)\\
&high&39.55&81.7(32.9,142.4)&38.1(30.5,51.4)\\
\hline
\multirow{3}{4em}{$Q$}
&low &0.04-0.07&0.27(0.02, 0.41)&0.20(0.15,0.27) \\
&medium &0.23-0.27&0.50(0.02,0.97)&0.15(0.10,0.21)\\
&high &0.47-0.77&0.59(0.03,0.98)&0.30(0.23,0.45)\\
\hline
\hline
\multirow{3}{4em}{D=G+P}
& low&&129.4=88.8+40.6  & 208.0=4.3+203.7\\
&medium&&9.8=0.52+9.2&77.7=0.20+77.1\\
&high&&7.5=1.0+6.5&38.2=0.1+38.1\\
\hline
\multirow{3}{4em}{MSE}
& low&&0.01 &0.02\\
&medium&&0.02&0.02\\
&high&&0.03&0.02\\
 \hline
 \hline
\end{tabular}
\end{center}
\label{table: Model1 post data}
\end{table}

\subsubsection{Two-zone model}
The near field box of size $(0.51 \text{m} \times 0.51 \text{m} \times 0.41 \text{m}=0.105\text{m}^3)$ was constructed within the far field box \citep{arnold}. The volume of the far field is $11.79$ m$^3$, which is the chamber volume minus the near field volume. The airflow parameter $\beta$ cannot be directly measured, but it was estimated from the local air speed to range from $0.24$ to $1.24$ m$^3$/min. 
Similar to the one-zone model, three different experimental data sets at three different ventilation levels were used. Table~\ref{table: Model2 post data} shows the medians and 95$\%$ C.I.s of the MCMC posterior samples, MSE and D=G+P. At all ventilation rates, non-Gaussian SSM 95$\%$ C.I.s include the true values of $Q$ but only at a medium ventilation rate, it includes the true value for $G$. The Gaussian SSM 95$\%$ C.I.s cover the true value of $Q$ at medium ventilation level but none of the generation rates $G$. The BNLR 95$\%$ C.I.s only cover the true value of $Q$ at a high ventilation level. The true value for $\beta$ was not directly measured and hence is unknown, however, it was estimated to be between 0.24 and 1.24. In general, non-Gaussian SSM 95$\%$ C.I.s for $\beta$ are closer to those values.

\begin{figure}[h!]
\centering
\captionsetup{justification=centering}
\hspace*{-0.35in}
\includegraphics[scale=0.8]{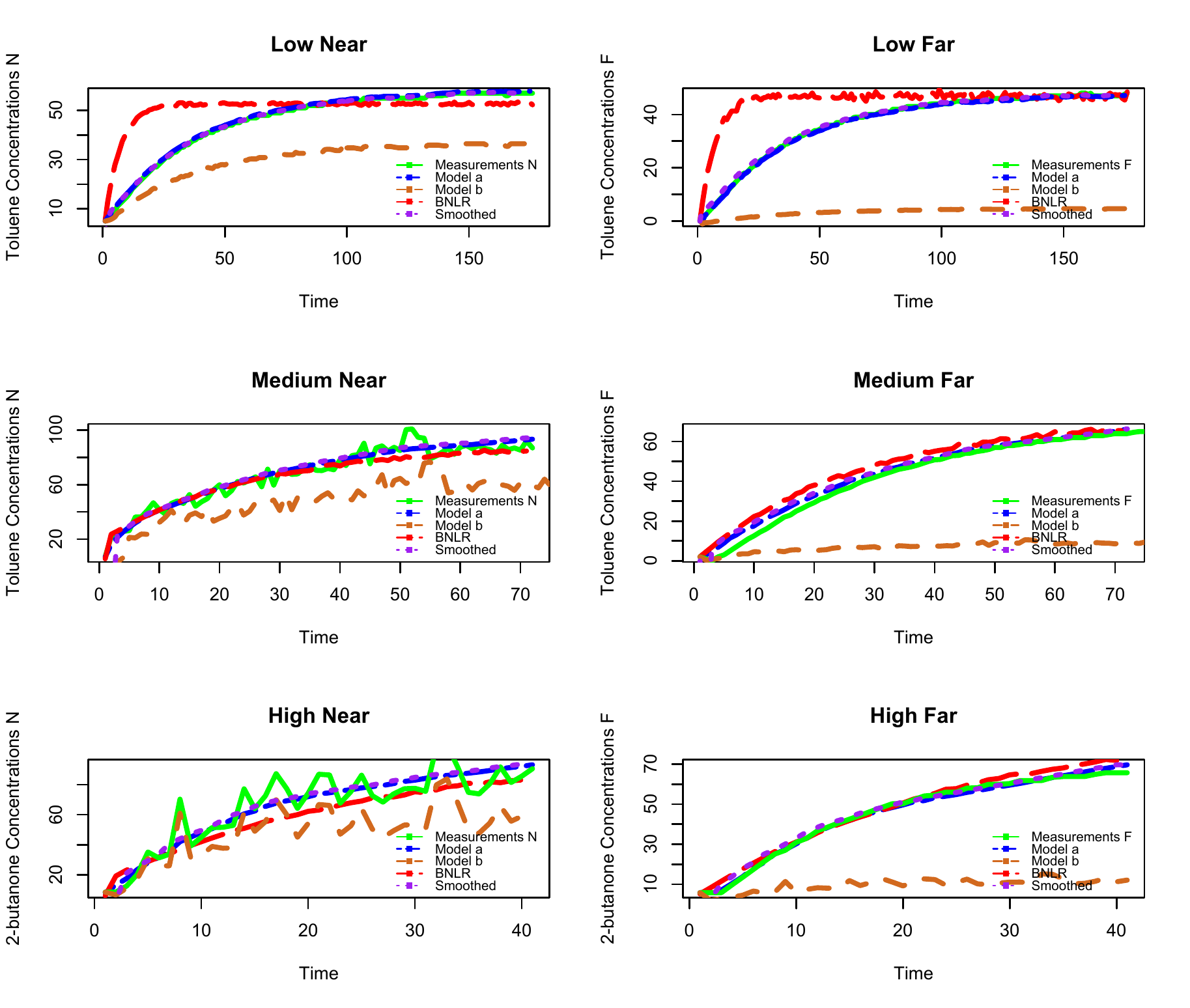}
 \caption{Plot of the measured concentrations and the mean of the posterior samples of the latent states conditional on the measurements in the near field and far field for:\\ a: Non-Gaussian SSM, b: Gaussian SSM and BNLR} 
 \label{fig: Model2 post data} 
 \end{figure}

MSE and D=G+P scores clearly indicate that non-Gaussian SSM produced better fit than the BNLR and the Gaussian SSM which is also confirmed in Figure~\ref{fig: Model2 post data}.
\begin{table}[h!]
\small
\caption{Posterior predictive loss (D=G+P), MSE, medians and 95\% C.I. of the posterior samples of the two-zone model parameters using toluene and acetone solvents}
\begin{center}
\setlength\tabcolsep{7pt}
\begin{tabular}{lcccccc}
  \hline 
\hline
Parameter &Ventilation  &True & Non-Gaussian   & Gaussian  &BNLR \\
 &level&value&SSM&SSM&\\
\hline
\multirow{3}{4em}{$G$} 
&low &43.2&30.4(30.0, 32.2)&115.8(88.9, 143.9)&28.1(28.0,28.4)\\
&med &86.4& 73.7(60.2,90.5)&141.6(130.6,149.7)&28.5(28.0,30.8)\\
&high &120.7&49.8(33.9,68.3)&132.9(121.6,148.0)&43.7(37.8,50.3)\\
\hline
\multirow{3}{4em}{$Q$}
&low&0.04-0.07&0.68(0.09, 0.98)&0.28(0.23,0.36)&0.62(0.60,0.65)\\
&med&0.23-0.27&0.38(0.11,0.50)&0.25(0.20,0.31)&0.38(0.29,0.50)\\
&high&0.47-0.77&0.46(0.45,0.98)&0.14(0.11,0.16)&0.5(0.30,0.64)\\
\hline
\multirow{3}{4em}{$\beta$}
&low&0.24-1.24&3.0(2.3,3.7)& 5.1(4.1,6.0)&4.9(4.7,5.0)\\
&med&0.24-1.24&2.9(2.5, 3.4)&2.3(2.0,2.8)&4.5(3.4,5.0)\\
&high&0.24-1.24&2.2(1.5, 2.8)&2.5(2.0,3.0)&4.1(2.7,4.9)\\
\hline
\hline
\multirow{6}{4em}{D=G+P}
& low&&5653=&	554650=&	248358=\\
&	&&189+5464&554234+416&73006+ 175352\\
&medium&&22262=&	850014=&	93267=\\
&	&&10596+11666&424452+425562&16824+76443\\
&high&&20941=&	479098=&	119212=\\
&	&&4345+16596&240278+238820&64968+54244\\
\hline
\multirow{3}{4em}{MSE}
& low&&0.62 &1835.2&129.2\\
&medium&&13.0&2952.4&96.5\\
&high&&52.9&2930.2&632.3\\
 \hline
   \hline
\end{tabular}
\end{center}
\label{table: Model2 post data}
\end{table}
 
\subsubsection{Turbulent eddy diffusion model} 
\cite{eddyram} constructed a chamber of size $(2.8 \text{m} \times 2.15 \text{m} \times 2.0 \text{m}=11.9\text{m}^3)$, where toluene was released. Measurements were taken at two locations at distances $0.41$ m and $1.07$ m away from the source every two minutes. Due to the limited spatial information from the two locations, an unstructured covariance for $\nu_t(s)$ was used instead of the geostatistical exponential covariance that was considered in the simulation analysis. Non informative prior was assigned to the covariance matrix using $IW(3,I)$ \citep{gelman}.    

Table~\ref{table: Model3 post data} shows the medians and 95$\%$ C.I.s of the MCMC posterior samples, MSE and the D=G+P. The value of $D_T$ is difficult to measure; hence, the true value is unknown. However, \cite{eddyram} demonstrated that most of the reported values of $D_T$ in literature range from 0.001 to 0.01 m$^2$/sec. The 95$\%$C.I.s for $D_T$ in non-Gaussian SSM lie within that range. In addition, the 95$\%$C.I.s of $G$ include the true value. The 95$\%$C.I.s of the Gaussian SSM do not include any of the true parameter values. Figure~\ref{fig: Model2 post data} shows that the latent state estimates for both models are closer to the measurements in the first location than in the second location. MSE and D=G+P scores show that non-Gaussian SSM provides a better fit.
\begin{table}[h!]
\small
\caption{Posterior predictive loss (D=G+P), MSE, medians and 95\% C.I. of the posterior samples of the turbulent eddy diffusion model parameters using toluene solvent}
\begin{center}
\setlength\tabcolsep{7pt} 
\setlength\belowcaptionskip{-20pt} 
\begin{tabular}{lcccc}
  \hline 
\hline
Parameter &True value& Non-Gaussian SSM & Gaussian SSM  \\
\hline
$G$&1318.33 &1207.3(1107.2,1371.7)& 1118.7(1104.5,1294.3)\\
$D_T$ &0.001-0.01&0.007(0.006,0.008)&0.67(0.64,0.78)\\
\hline
\hline
D=G+P&&100877.8=59369.9+41507.9&32383410=258952.4+32124457\\
\hline
MSE&&337.3&1454.8\\
 \hline
   \hline
\end{tabular}
\end{center}
\label{table: Model3 post data}
\end{table}
\begin{figure} [h!]
\centering
\captionsetup{justification=centering}
\hspace*{-0.35in}
\includegraphics[scale=0.6]{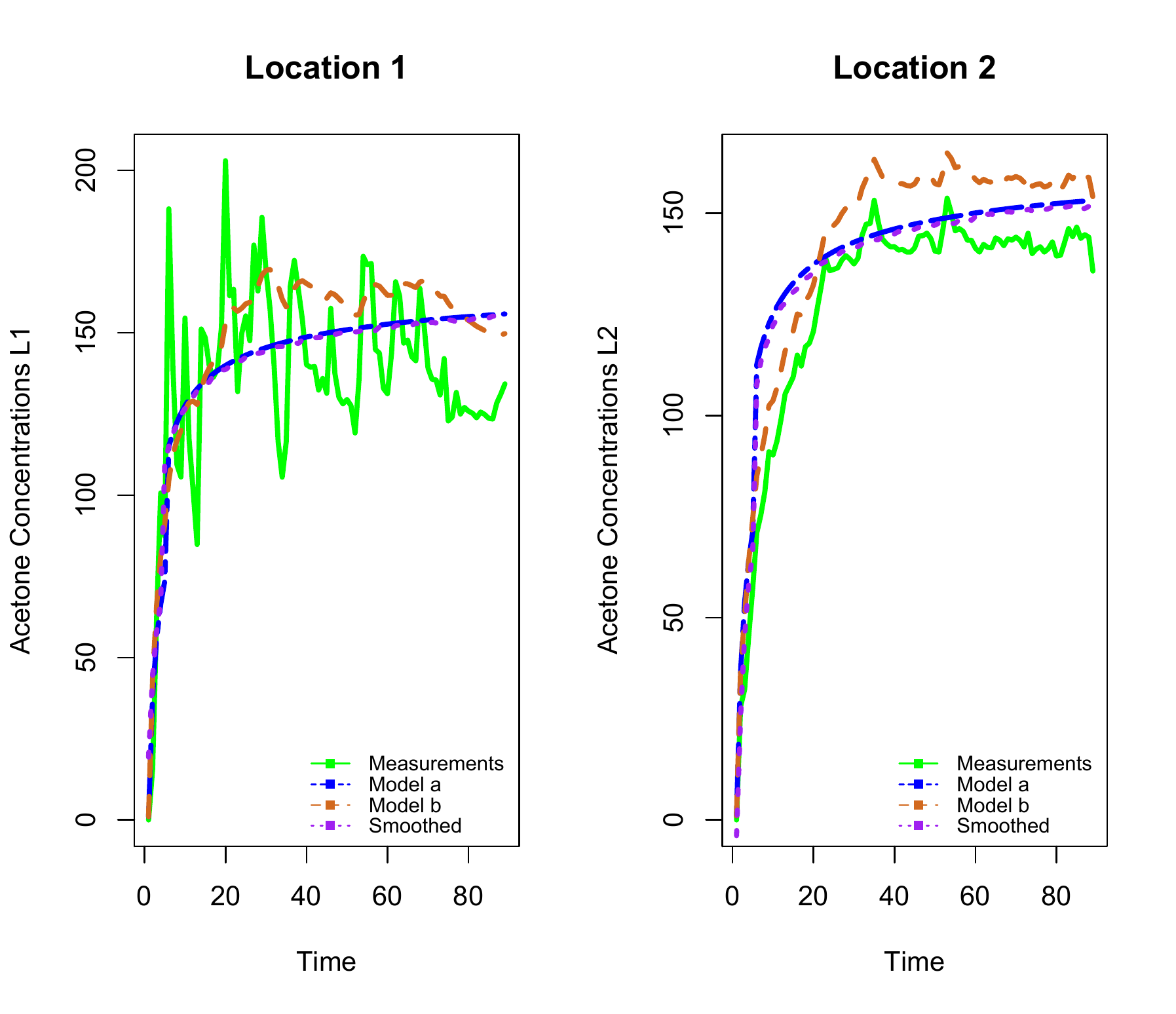}
 \caption{Plot of the measured concentrations and the mean of the posterior samples of the latent states conditional on the measurements at the two locations for:\\ a: Non-Gaussian SSM and b: Gaussian SSM} 
 \label{fig: Model3 post data} 
 \end{figure}

\section{Discussion}\label{sec:discussion}
We have proposed a framework of Bayesian SSMs for analyzing experimental exposure data specific to industrial hygiene. This approach combines information from physical models of industrial hygiene, observed data and prior information. We derive a likelihood by discretizing the physical models. It also expands upon the Gaussian noise assumptions, hence industrial hygienists will not be restricted to Gaussian SSMs.

In practical industrial hygiene settings, Gaussian SSMs are still often used as approximations to analyze possibly non-Gaussian data. To do so, some possibly inappropriate accommodations may need to be made. For example, \cite{pm10} allowed negative values in estimating PM$_10$ concentrations, while \cite{kfgas} used Kalman filters to predict gas concentrations by using a tuning parameter to fix $\sigma^2_\omega$ and $\sigma^2_\nu$ in a one dimensional autoregressive exposure model, rather than pursuing full statistical inference. Our simulation experiments and results demonstrate that Gaussian SSM's may yield extremely poor fits when data are non-Gaussian.  This was especially evident for the two-zone analysis. Our results will, we hope, inform the industrial hygiene community about some of the pitfalls of Gaussian SSMs.   

Non-Gaussian SSM's tended to perform better than linear Gaussian SSM's, a result that appeared to be consistent across different exposure models and different experimental conditions.  Moreover, our analysis of the two-zone data revealed that the discretized models outperform the BNLR method proposed by \cite{bayesih2z} for two zone data. This is unsurprising given that our approach is richer by accommodating stochastic distributions at two levels---one each for the measurement and transition equations---whereas BNLR accommodates only an error distribution from a nonlinear regression. Finally, our proposed approach also enjoys better interpretation than the hierarchical Gaussian process models of \cite{bayesih} as they provide greater precisions in estimates because the random effects in the hierarchical models of \cite{bayesih} tend to inflate variances.  

The eddy diffusion data has some limitations related to the small size of the chamber, which rendered a small difference between the concentrations in the two locations which also makes it hard to measure the spatial variation for Model~(\ref{eq: eddy_discrete}) implementation. Despite that, in most cases, a nonlinear non-Gaussian Bayesian SSM was able to characterize the data well and the model seems robust to most of the experimental scenarios. 

We conclude with some indicators for future research. First, as alluded to earlier, we will need to do a much more comprehensive spatiotemporal analysis for eddy diffusion experiments. While our simulation experiments showed the promise of spatiotemporal SSM's in analyzing eddy diffusion experiments, our chamber data analysis had limited scope because of the very small number of spatial measurements. Another important consideration is misaligned data, such as was considered in \cite{bayesih} for two zone experiments where not all measurements for the near and far fields came from the same set of timepoints. An advantage of the Bayesian paradigm is that we can handle missing data, hence misaligned data, very easily and indeed our Bayesian SSMs should be able to handle them as easily as the models in \cite{bayesih}. Future work will include such analysis and also extensions to spatiotemporal misalignment for eddy-diffusion experiments, where not all timepoints generated measurements for the same set of spatial locations. 


\newpage
 \bigskip
\begin{center}
{\large\bf SUPPLEMENTARY MATERIAL}
\end{center}

\begin{description}
\item[R-code for Bayesian SSMs used:] R- code to perform the filtering, smoothing and parameters estimation and model assessment methods described in the article. (BSTSP Rmd file)
\item[Discretization of the differential equations:] We approximate the deterministic physical model through discretization. The Taylor expansion of $C(t)$ at $t=t^*$ is $C(t)=\sum_{n=0}^{\infty} \frac{C^{(n)}(t^*)}{n \,!}(t-t^*)^n$, where $C^{(n)}(t^*)=\frac{d^n}{dt^n}C(t) \Bigr|_{t=t^*}$. 
Let $t=t^*+\delta_t$ hence \begin{equation} C(t^*+\delta_t)=\sum_{n=0}^{\infty} \frac{C^{(n)}(t^*)}{n \,!}(\delta_t)^n=C(t^*)+\frac{C'(t^*)}{1 \,!}\delta_t+o(\delta_t), \end{equation} for small $\delta_t$. From the above equation we can express $C'(t^*)$ as
 \begin{align}\label{eq:diffapprox} C'(t^*)=\frac{C(t^*+\delta_t)-C(t^*)}{\delta_t} +o(\delta_t).\end{align}
 
In the applications to the three physical models we replace the first order derivative $\frac{d}{dt}C(t)$ at $t=t^*$ with equation (\ref{eq:diffapprox}) using the appropriate value of $\delta_t$. In the one zone and two-zone models a value $\delta_t=0.01$ was found to provide an accurate approximation, while for the eddy diffusion model $\delta_t=1$ was used.

\item[Steady states derivations:] The steady state is achieved as $t\rightarrow \infty$ in the exact solution of the ODE. 
\begin{align}\label{eq:limit}
\lim_{t \to \infty} \text{exp}\{tF_t\}C(t_0)+F_t^{-1}[\text{exp}\{tF_t\}-I]g.\end{align}\\
For the one zone model $F_t=-(Q+K_LV)/V$ and $g=G/V$ so $\ref{eq:limit}=
F_t^{-1}[-I]g=G/(Q+K_LV)$. Since $K_L$ is usually small, it can be approximated by G/Q. Hence as $t\rightarrow \infty$ $C(t)\approx G/Q$.

For the two zone model, $F_t=A=\left[{\begin{array}{cc} -\beta/V_N & \beta/V_N \\
\beta/V_F & -(\beta+Q)/V_F +K_L \end{array} } \right] $ and $g=\left[{\begin{array}{c} G/V_N\\ 0 \end{array} }\right]$. Since $K_L$ is usually small it can be ignored for simplicity.
The term $\text{exp}(tF_t)$, where $\text{exp()}$ is the matrix exponential, can be written as $\text{exp}(tL \Lambda L^{-1})=\sum e^{t\lambda}G_i$ where $G_i = u_i v_i^{T}$, $u_i$ is the $i$-th column of $L$ and $v_i^{T}$ is the $i$-th row of $L^{-1}$. It easily follows that $e^{tF_t} = \sum\limits_{i=1}^m e^{t\lambda_i}G_i$. The eigenvalues are available in closed form \cite{bayesih2z} as
\begin{equation}
\begin{array}{l}
\lambda_1 = \frac{1}{2}\left[-\left(\frac{\beta V_F+(\beta+Q)V_N}{V_N V_F}\right)+\sqrt{\left(\frac{\beta V_F+(\beta+Q)V_N}{V_N V_F}\right)^2-4\left(\frac{\beta Q}{V_N V_F}\right)}\right],\\
\\
\lambda_2= \frac{1}{2}\left[-\left(\frac{\beta V_F+(\beta+Q)V_N}{V_N V_F}\right)-\sqrt{\left(\frac{\beta V_F+(\beta+Q)V_N}{V_N V_F}\right)^2-4\left(\frac{\beta Q}{V_N V_F}\right)}\right].\\
\end{array}\label{eigenvalues}
\end{equation}
As long as $\beta$ and $Q$ are positive, the sum of the two eigenvalues are negative. Hence $e^{tF_t} = \sum\limits_{i=1}^m e^{t\lambda_i}G_i \rightarrow 0$ as $t\rightarrow \infty$ and the first term becomes $0$ and the second term becomes $A^{-1}[-I]g$.
The determinant of $A$ is det$(
{A})=Q \beta/V_N V_F$, and $A^{-1}=\left[{\begin{array}{cc} -((\beta+Q)/V_F)(V_NV_F/\beta Q) & -(\beta/V_N)(V_NV_F/\beta Q) \\
-(\beta/V_F)(V_NV_F/\beta Q) & -((\beta)/V_N )(V_NV_F/\beta Q)\end{array} } \right] $.
So the steady state is a $2 \times 1$ vector equal to $A^{-1}[-I]g=\left[{\begin{array}{c}\frac{G}{Q}+\frac{G}{\beta} \\ \frac{G}{Q}\end{array}}\right]$. So as $t\rightarrow \infty$ $C_N(t)\approx \frac{G}{Q}+\frac{G}{\beta}$ and $C_F(t)\approx \frac{G}{Q}$.

The steady state for the eddy diffusion model is theoretically the value of $C(s,t)$ in equation (\ref{eq:eddyexact}) when $t\rightarrow \infty$. Clearly $\lim_{t \to \infty} \frac{G}{2 \pi D_T (||s||)}\left(1- erf \frac{||s||}{\sqrt{4D_T t}}\right)=\frac{G}{2 \pi D_T (||s||)}$.

\end{description}
 \clearpage
\bibliography{mybib}
\bibliographystyle{plain}
\end{document}